\documentclass [aps,prb,twocolumn,amssymb,showpacs,superscriptaddress,reprint,citeautoscript]{revtex4-1}
\usepackage[pdftex]{graphicx}%

\begin{document}

\author{R. T. Gordon}
\email{rt-gordon@wiu.edu}
\affiliation{Department of Physics, Western Illinois University, Macomb, IL 61455, USA}
\altaffiliation{work done while at The Ames Laboratory and Department of Physics \& Astronomy, Iowa State University, Ames, IA 50011, USA}

\author{N. D. Zhigadlo}
\email{zhigadlo@phys.ethz.ch}
\affiliation{Laboratory for Solid State Physics, ETH Zurich, CH-8093 Zurich, Switzerland}

\author{S.~Weyeneth}
\email{wstephen@physik.uzh.ch}
\affiliation{Physik-Institut der Universit\"at Z\"urich, Winterthurerstrasse 190, CH-8057 Z\"urich, Switzerland}

\author{S. Katrych}
\email{katrych@phys.ethz.ch}
\affiliation{Institute de Physique de la Mati\`ere Complexe, Ecole Polytechnique F\'ed\'erale de Lausanne (EPFL), CH-1015 Lausanne, Switzerland}
\altaffiliation{work done while at Laboratory for Solid State Physics, ETH Zurich, CH-8093 Zurich, Switzerland}

\author{R. Prozorov}
\email[corresponding author: ]{prozorov@ameslab.gov}
\affiliation{The Ames Laboratory and Department of Physics \& Astronomy, Iowa State University, Ames, IA 50011, USA}

\title{Conventional superconductivity and hysteretic Campbell penetration depth in single crystals MgCNi$_3$}

 \date{19 February 2013}
 
 \begin{abstract}
 Single crystals of MgCNi$_3$, with areas sized up to 1 mm$^{2}$, were grown by the self flux method using a cubic anvil high pressure technique. In low applied fields, the \textit{dc} magnetization exhibited a very narrow transition into the superconducting state, demonstrating good quality of the grown crystals. The first critical field \textit{H$_{c1}$}, determined from a zero temperature extrapolation, is around 18 mT. Using the tunnel - diode resonator technique, the London penetration depth was measured with no applied \textit{dc} field and the Campbell penetration depth was measured with the external \textit{dc} fields up to 9~T for two different sample orientations with respect to the direction of applied magnetic field. The absolute value of the London penetration depth, $\lambda(0) = 245 \pm 10$~nm was determined from the thermodynamic Rutgers formula. The superfluid density, $\rho_s=(\lambda(0)/\lambda(T))^2$ was found to follow the clean isotropic \textit{s}-wave behavior predicted by the weak - coupling BCS theory in the whole temperature range. The low - temperature behavior of the London penetration depth fits the BCS analytic form as well and produces close to the weak - coupling value of $\Delta (0)/k_BT_c = 1.71$. The temperature dependence of the upper critical field, \textit{$H_{c2}$}, was found to be isotropic with a slope at \textit{T$_c$} of -2.63 T/K and \textit{H$_{c2}$}(0) $\approx$ 12.3 T at zero temperature. The Campbell penetration depth probes the vortex lattice response in the mixed state and is sensitive to the details of the pinning potential. For MgCNi$_3$, an irreversible feature has been observed in the TDR response when the sample is field-cooled and warmed versus zero-field-cooled and warmed. This feature possesses a non-monotonic field dependence and has commonly been referred to as the peak effect and is most likely related to a field - dependent non - parabolic pinning potential.
\end{abstract}

 
 
 \maketitle

	\begin{figure*}[tb]
	\begin{center}
	\includegraphics[width=17cm]{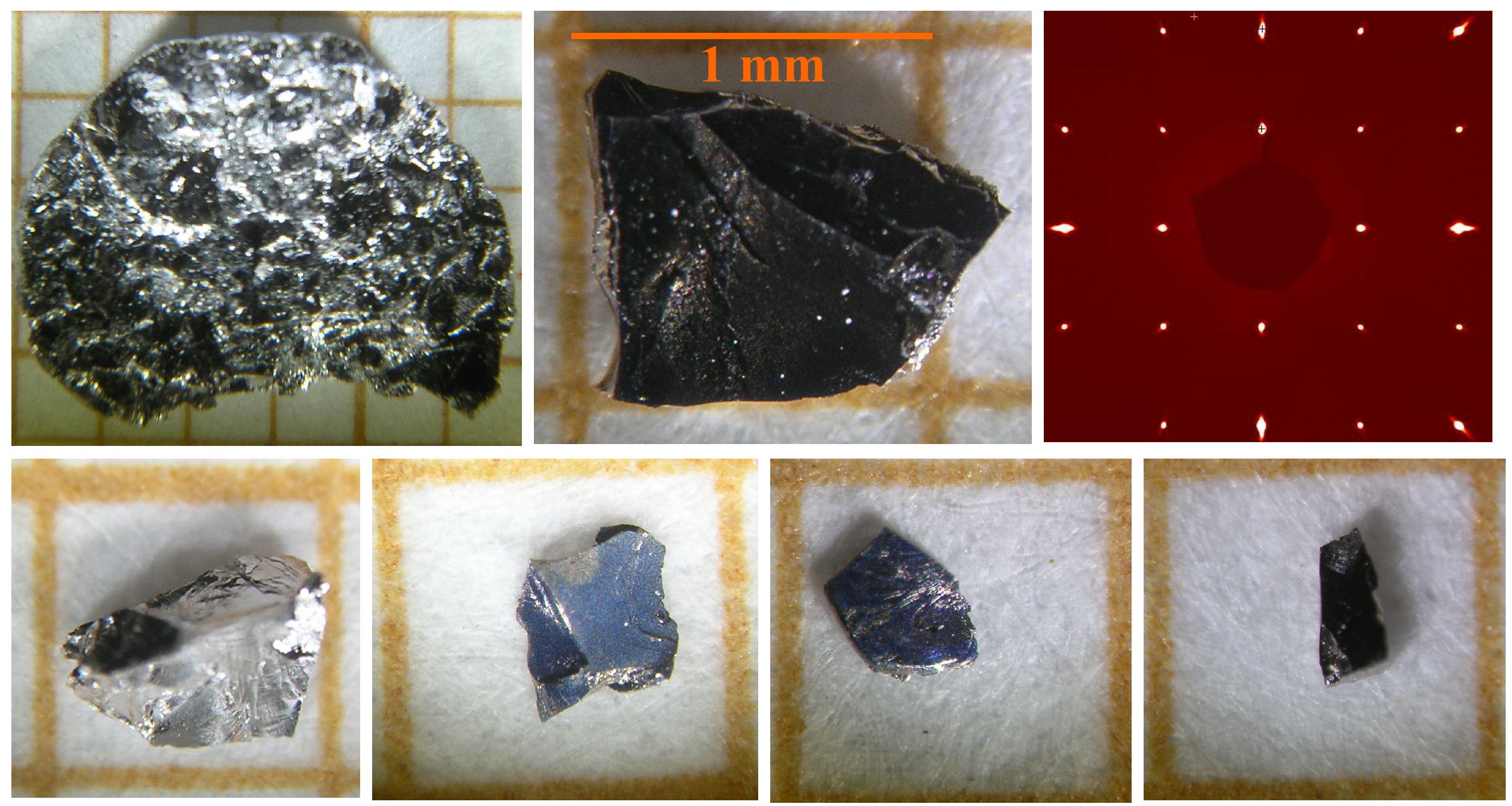}
	\caption{(Color online) Optical microscope images of MgCNi$_3$ single crystals. An as grown melted lump with a mixture of single-crystalline MgCNi$_3$ and some fluxes are shown in the left upper corner. After crushing the lump, a large number of crystals with the sizes up to 1 mm$^2$ were found. Upper right frame shows the \textit{hk0} reciprocal space section determined by XRD of single crystal MgC$_{0.92}$Ni$_{2.88}$.}
	\label{figNZ1}
	\end{center}
	\end{figure*}

   The announcement of superconductivity in the intermetallic compound MgCNi$_3$ has generated a great amount of excitement since its discovery in 2001 \cite{He2001}.  This material has gained so much interest because it is a superconductor with a transition temperature near 7 K and it shares the same perovskite structure as that of the high-$T_c$ cuprates but with the O atoms replaced by Ni.  After the realization of these facts, many began to consider the possibility that this material could bridge the gap between conventional superconductivity in intermetallic compounds and unconventional superconductivity in high-$T_c$ superconducting oxide perovskites. This material has also been described as being a three dimensional analog to the two dimensional family of borocarbide superconductors.  A possible scenario, brought about by predictions made for high-$T_c$ superconductors, is that the superconducting state arises due to interactions involving ferromagnetic spin fluctuations from the large Ni concentration.  Band structure calculations indicate that the abundance of Ni in this compound places it near a ferromagnetic instability \cite{Shim2001, Singh2001} and the existence of such a peak is confirmed by both photoemission and x-ray absorption experiments \cite{jhkim2002}.

      Although some have made claims of observations of unconventional superconducting properties, there are many measurements which indicate usual BCS behavior. C$^{13}$ NMR investigations have found that the nuclear spin-lattice relaxation rate $1/^{13}T_1$ exhibits the typical exponential behavior expected for $s$-wave superconductivity \cite{Singer2001}.  One point contact spectroscopy study concludes conventional weak - coupling BCS \textit{s}-wave superconductivity \cite{Shan1_2003}, while another tunneling junction measurements have inferred the magnitude of the superconducting gap from an observed zero bias conductance peak and have found it to be larger than that of the weak coupling BCS value, which has led them to make the conclusion that the electron coupling in this material is strong \cite{Mao2002}.  A carbon isotope effect has been observed in this compound, indicating that the carbon-based phonons do play an important role in the superconductivity and is thus in support of the conventional BCS phonon mediated model of superconductivity \cite{Klimczuk2004}.  Muon spin rotation studies have found evidence for BCS behavior in the superconducting gap \cite{Macdougall2006}.  Measurements of the specific heat are indicative of a fully gapped superconducting state but they do not seem to be in agreement on the strengh of the coupling or the effects of spin fluctuations \cite{Lin2003, Shan2003}.  Electrical transport measurements have found that the normal state resistivity follows a conventional electron-phonon scattering model and that $H_{c2}$ near $T_c$ is linear.  They have used these findings to conclude that MgCNi$_3$ is a conventional BCS superconductor \cite{Lee2008}.  Previous tunnel diode resonator experiments on this material have been performed on polycrystalline samples and powders and it was found that the low temperature behavior was quadratic \cite{Prozorov2003}, which would point to the existence of nodes in the superconducting gap function. However, it may have been the case that inter-grain interactions in those samples studied may have influenced the data.  Most of the scattered physical properties and theoretical calculations for MgCNi$_3$ are reviewed in Ref.~\cite{Mollah2004}. Later, with the appearance of the first single crystal data \cite{Lee2008,Lee2007,Pribulova2011,Diener2009}, it became clear that there were some contradictions regarding the physical property measurements obtained on polycrystalline and single crystalline samples. In addition, the recently observed peak effect and dynamics of vortex matter in MgCNi$_3$ \cite{Jang2009} requires further detailed investigations. In this sense, MgCNi$_3$ single crystals, which obey a simple perovskite cubic crystal structure, provide an interesting possibility for further magnetic studies.

In this paper, high-pressure crystal growth and precision measurements of the magnetic penetration depth on bar shaped MgCNi$_3$ crystals for two different sample orientations in fields from 0 to 9 T are reported.  The superfluid density is constructed from the zero field penetration depth and this data has been shown to agree well with the isotropic BCS \textit{s}-wave superfluid density model.  The \textit{H}$_{c2}$(T) curve is also constructed for fields applied in two different directions and it is found to be linear near $T_c$ and also isotropic, indicating that the change in \textit{T}$_c$ due to demagnetization effects from the sample shape are negligible.  The penetration depth in field, consisting of London and Campbell components, shows a very interesting hysteresis when the sample is zero-field-cooled, field-warmed and then field-cooled, most likely due to a vortex lattice response referred to as the peak effect.

Due to the high volatility of Mg and the relatively poor reactivity of C, it is extremely difficult to synthesize single phase samples of MgCNi$_3$, even in polycrystalline form. The synthesis of single crystals is not possible in an open system; however, it can be done under high pressure, as was first demonstrated by Lee {\it et al.} \cite{Lee2007}. As both methods show, the superconductivity in this material is very sensitive to the details of heat treatment and final stoichiometry. Amos {\it et al.} \cite{Amos2002} reported that different C contents in MgC$_x$Ni$_3$ polycrystals caused different cubic cell parameters: \textit{a} increased from 3.795 to 3.812~\AA\ as $x$ varied from 0.887 to 0.978. In addition, the superconducting transition temperature $T_c$ sensitively depends on the real C content and decreases with increasing content. In contrast to polycrystalline MgCNi$_3$, the Ni site was partly deficient in single crystals synthesized under high pressure conditions \cite{Lee2007}.

Here we report our successful growth process for MgCNi$_3$ single crystals together with their structural and superconducting properties. The single crystals of MgCNi$_3$ were grown at ETH Zurich using cubic anvil high-pressure and high-temperature techniques. The mixture of Mg, C, and Ni powders in a molar ratio 1:1:3 were placed inside of a BN crucible with the inner diameter of 6.8 mm, and the length of 8.5 mm. The heating element is a graphite tube. Six anvils generate pressure on the whole assembly. In a typical run, a pressure of 3 GPa is applied at room temperature. While keeping pressure constant, the temperature is ramped up within 2 h to the maximum value of 1600-1700 $^o$C, and is kept stable for 1 h and then slowly cooled to room temperature. The high pressure was maintained constant throughout the growth and was removed only after the end of the crystal growth process. The final product was a melted lump with a mixture of single-crystalline MgCNi$_3$ and some fluxes (see left upper corner image in Fig.~\ref{figNZ1}). After crushing the lump, the single crystals with various shapes and of sizes up to 1 mm$^2$ were mechanically extracted (Fig.~\ref{figNZ1}).

	\begin{figure}[tb]
	\begin{center}
	\includegraphics[width=9cm]{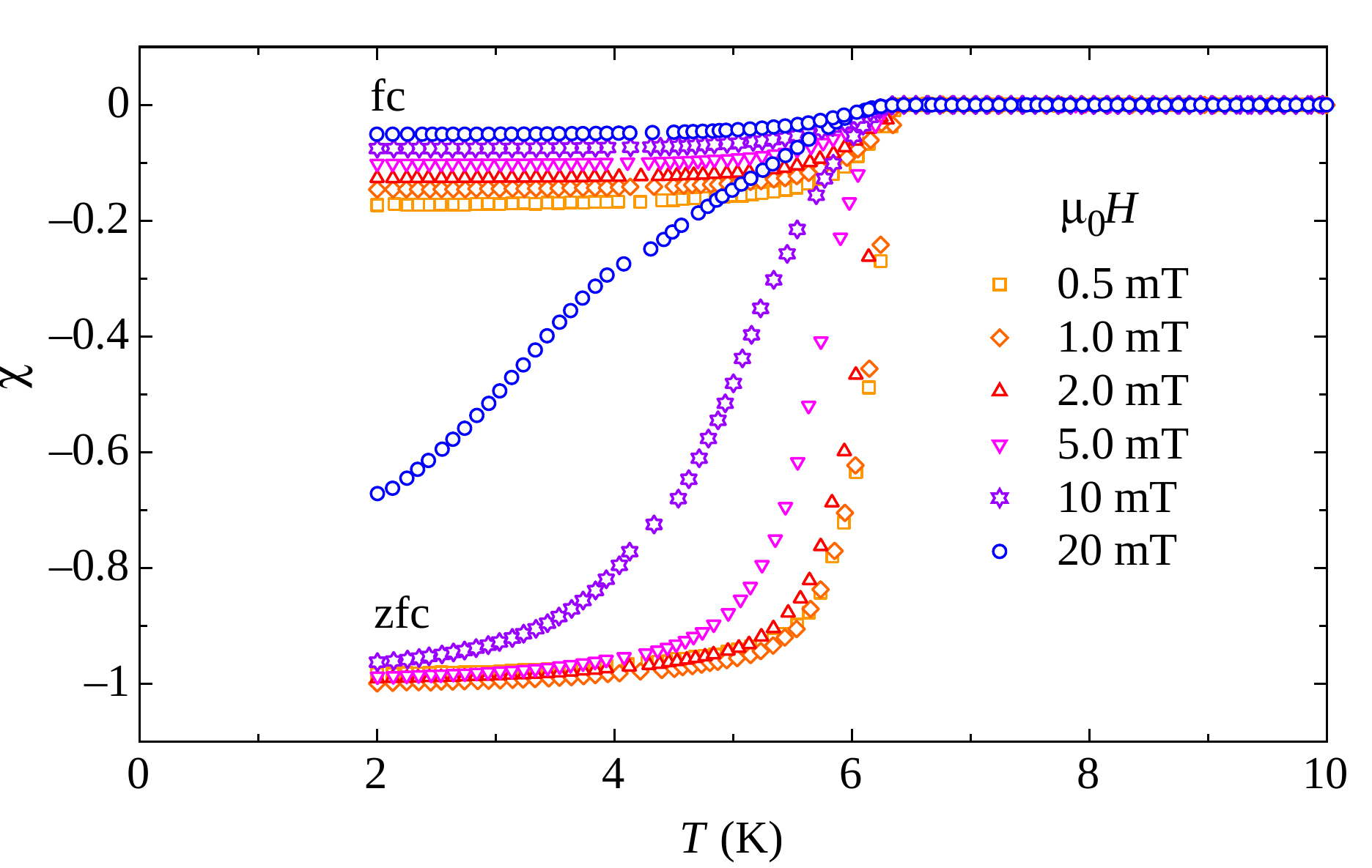}
	\caption{(Color online) Magnetic susceptibility $\chi$ as a function of temperature $T$ for various magnetic fields $H$. }
	\label{SW1}
	\end{center}
	\end{figure}
	
The quality of the crystals was checked by using a single crystal x-ray diffractometer equipped with a CCD area detector (Xcalibur PX, Oxford Diffraction), which allowed us to examine the whole reciprocal space (Ewald sphere) for the presence of other phases or crystallites with different orientations. As it is clearly seen in the upper right frame of Fig.~\ref{figNZ1}, no additional phases, impurities, or intergrowing crystals were detected by examination of the reconstructed reciprocal space. The crystal structure was determined by a direct method and refined on \textit{F$^2$}, employing the programs SHELXS-97 and SHELXL-97 \cite{Sheldrick}. All atomic positions were found by a direct method. After several refinement cycles the correct crystallographic composition was determined and the final R factor was 1.8\% indicating the high quality of the structural model. The occupation parameters for the Mg, C, and Ni were found to be 1: 0.92, and 2.88, respectively. Thus, according to the structural analysis, the more appropriate chemical formula for our crystals is MgC$_{0.92}$Ni$_{2.88}$. Single crystal analysis confirmed the cubic structure with lattice parameter $a=3.7913(1)$~\AA. This value of lattice constant a is slightly smaller than that of observed in MgCNi$_{2.8}$ ($a=3.812$~\AA, $T_c=6.7$~K) single crystals grown at the pressure of 4.25 GPa and temperature 1200 $^o$C \cite{Lee2007}. Lee \textit{et al.} also note that crystals grown under a pressure below 3.5 GPa had C deficiencies. \cite{Lee2007} The present data confirm this observation. However, in our high pressure growth conditions, besides the C deficiency, the Ni site was also partially deficient and thus the resulting $T_c$ is reduced more. For various growth batches, $T_c$ varies between 6.4 and 6.8 K.  For measurements of superconducting properties, clean, flat single crystals with sizes of a few hundred micrometers were selected.
	
		\begin{figure}[tb]
		\begin{center}
		\includegraphics[width=9cm]{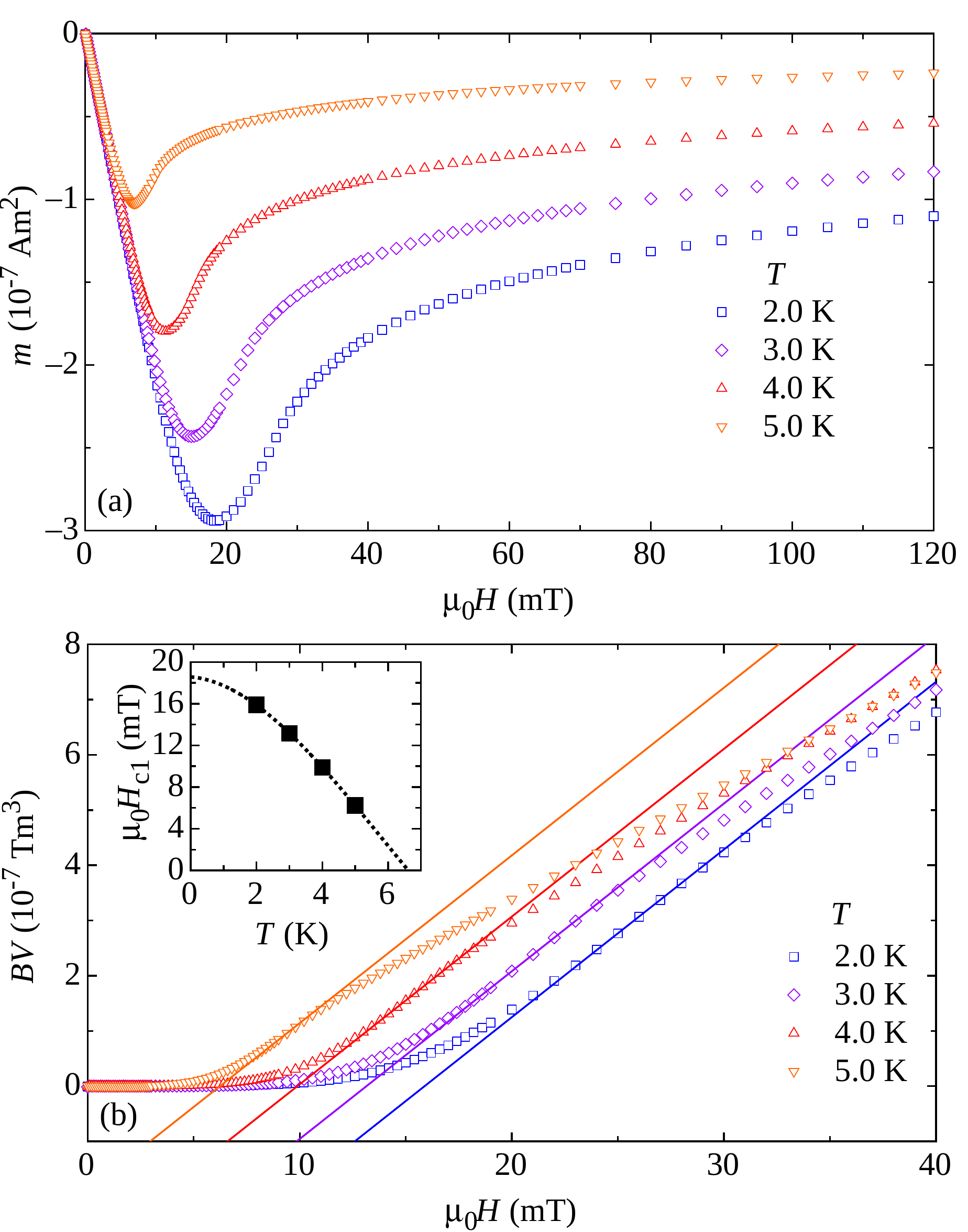}
		\caption{(Color online) Analysis in order to extract the lower critical field $H_{c1}$ from $m(H)$ measurements of the studied MgCNi$_3$ single crystal. Upper panel: As measured magnetic moment $m$ versus the applied magnetic field $H$. Lower panel: Magnetic induction $BV$ versus the applied magnetic field along the $ab$-plane. The inset presents the determined $H_{c1}$ as a function of temperature. We estimate the zero temperature value at $\mu_0H_{c1}(0)\simeq18$~mT.}
		\label{SW2}
		\end{center}
		\end{figure}
	
The susceptibility $\chi$ of a flat, plate-like single crystal with approximate dimensions $0.8\times0.4\times0.1$~mm$^3$ was measured using a Quantum Design MPMS magnetometer as a function of temperature for various magnetic fields applied along the planar sample. Field-cooled (fc) and zero-field-cooled (zfc) temperature dependent measurements in low fields are shown in Fig.~\ref{SW1}. The transition to the superconducting state in low fields is very narrow, demonstrating the good quality of the single crystal. Supplemental magnetization curves were recorded for this single crystal in the temperature range between 2 and 5 K. From such measurements, the lower critical field $H_{c1}$ was determined using a procedure discussed elsewhere \cite{NZ}. For this the magnetic induction $B$ was determined from the measured magnetic moment and plotted as a function of magnetic field (see Fig.~\ref{SW2}). Due to the uncertainty of the sample volume $V$, only the product $BV$ was calculated and plotted according to
\begin{equation}
B=\mu_0(M+H)=\mu_0(m/V+H)
\end{equation}
Since $B=0$ in the Meissner state it is possible to calculate from the data of $m(H)$, the field above which this equality is invalid. The sudden increase from zero occurs due to the penetration of vortices. The resulting $\mu_0H_{c1}(0)\simeq18$~mT is consistent with a magnetic penetration depth of $\sim200$~nm, invoking a $\kappa\simeq100$. All MgCNi$_3$ single crystals investigated in this work did not show any traces of ferromagnetism, in contrast to a recent report of ferromagnetic domains coexisting with superconductivity in carbon deficient  MgCNi$_3$.\cite{Kuniaz2012}

The single crystalline MgCNi$_3$ sample was studied using a tunnel diode resonator (TDR) circuit technique. A detailed description of the application of this technique to study London and Campbell penetration depths in superconductors can be found in Ref.~\onlinecite{Prozorov2006rev,Prozorov2011}.  The principle elements of the setup consist of an $LC$ self-oscillating circuit supported by a tunnel diode.  The tunnel diode has a heavily doped and extremely thin (10 nm) p-n junction, which gives it useful properties not common to ordinary diodes. The $IV$ curve contains a region of negative differential resistance and when the diode is biased to this region, it acts as an $AC$ power source for the tank circuit. The tank circuit oscillates with a natural resonance frequency of $f_0=1/2\pi\sqrt{LC}$, which is very near 14 MHz.  The sample to be studied is mounted on a sapphire rod and inserted into the inductor coil of the oscillator.  The sample changes the resonance frequency of the circuit through its interaction with the $ac$ magnetic field of the coil, which is on the order of 0.1 $\mu$T.  This small value of the excitation field of the coil ensures that its effect on the state of the sample is negligible and hence this technique is non-perturbative.  For a superconductor below its critical temperature, the $ac$ magnetic field of the coil has a characteristic decay length, commonly referred to as the London penetration depth $\lambda$, which is a function of temperature.  The measured change in frequency, $\Delta f$, is proportional to the dynamic magnetic susceptibility of the sample.  This susceptibility may be written in terms of this penetration depth and a characteristic radius of the sample $R$, which is calculated using a procedure given in reference \cite{Prozorov2000}.  So we have,

    \begin{equation}
\Delta f\left(  T\right)  =-G\chi\left(  T\right)  =G\left[
1-\frac{\lambda}{R}\tanh\left(  \frac{R}{\lambda}\right)  \right]
\label{df},
\end{equation}

    \noindent where the geometry dependent calibration factor is expressed as $G\simeq f_{0}V_{s}/2V_{c}\left(  1-N\right)$, $V_s$ is the sample volume, $V_c$ is the effective coil volume and \textit{N} is the demagnetization factor of the sample.  The factor \textit{G} can be measured directly by extracting the sample from the coil at the lowest temperature of the experiment.  Since the effective radius of the sample, $R$, is much greater than the penetration depth, $\lambda$, this expression can be rewritten so that changes in the resonant frequency are proportional to changes in the penetration depth

    \begin{equation}
    \Delta f\left( T\right) \propto \Delta \lambda \left( T\right).
    \end{equation}

    The most valuable feature that this technique has to offer is not the ability to measure the actual value of the penetration depth, but rather its variation with temperature to great precision, $\Delta\lambda=\lambda(T)-\lambda(T_{min})$, with $T_{min}$ being the minimum temperature that can be reached during the experiment.  The noise level of the system used for this experiment is $\approx$ 0.1 Hz/hour, which combined with the natural resonance frequency of the system of 14 MHz corresponds to a resolution on the order of parts per billion.  This level of precision allows for the measurement of $\Delta\lambda$ to a single \AA ngstr\"om. The circuit assembly is mounted inside of a $^3$He refrigerator that is lowered into the bore of a superconducting solenoid allowing for the application of $dc$ fields up to 9 T in addition to the extremely small $ac$ field supplied by the TDR.

\begin{figure}[htb]
\begin{center}
\includegraphics[width=9cm]{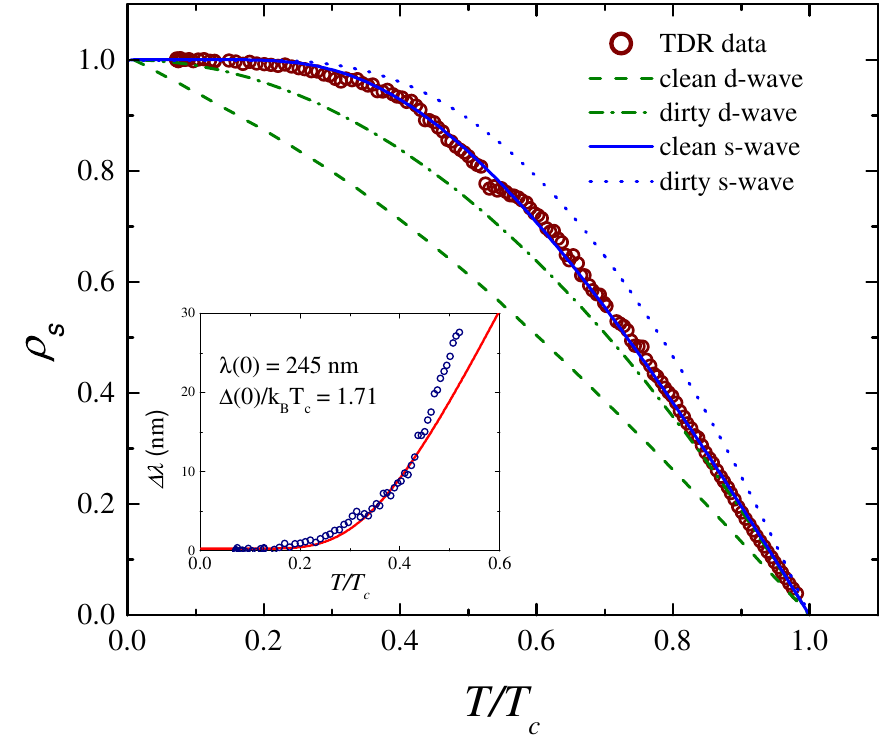}
\caption{(Color online) The superfluid density, $\rho_s(T)$, constructed from the London penetration depth measured by a tunnel diode resonator (red circles). Symbol size represents a $\pm 10$nm error bar. Expectations for the single - gap BCS superconductors are shown for clean - limit $s-$wave (solid blue), clean - limit $d-$wave (dashed green), dirty - limit $s-$wave (dot blue) and dirty - limit  $d-$wave (dashed - dot green). Inset shows low - temperature variation of the London penetration depth and a weak - coupling BCS isotropic s-wave fit with a fixed $\lambda(0) = 245$~nm and the gap, $\Delta (0)$, as a free parameter.}
\label{rhos}
\end{center}
\end{figure}

 The superfluid density, $\rho_s$, is an important quantity that can be related to the gap structure of a superconducting through the London penetration depth \cite{Prozorov2006rev,Prozorov2011}. If the zero-temperature value of the penetration depth, $\lambda(0)$, is known, then the superfluid density can be constructed from $\Delta\lambda$ as,

    \begin{equation}
    \rho_s(T)=\left(\frac{\lambda(0)}{\lambda(T)}\right)^2= \left[ 1+\frac{\Delta\lambda (T)}{\lambda\left( 0 \right)} \right]^{-2}
    \end{equation}

\noindent where $\Delta\lambda (T)$ is the measured variation of the London penetration depth, measured by using a TDR and by applying the calibration procedure described previously. Without a direct measurement, the most reliable procedure to evaluate $\lambda(0)$ is to use thermodynamic Rutgers formula that can be written as \cite{kim2013},

\begin{equation}
\left\vert \frac{d\rho }{dt}\right\vert _{T\rightarrow T_{c}}=\frac{16\pi ^{2}}{\Phi _{0}}\frac{\Delta C}{\left\vert \frac{dH_{c2}}{dT}\right\vert _{T\rightarrow T_{c}}}\lambda ^{2}\left( 0\right) 
\end{equation}

Taking the measured slope, $\vert dH_{c2}/dT \vert _{T \rightarrow T_c}=2.63$~T/K and the jump of electronic specific heat at $T_c$, $\Delta C = 129$~mJ/(mol(Ni) K) \cite{Shan2003}, and using the iterative procedure descrived in Ref.~\onlinecite{kim2013}, we obtain $\lambda(0) = 245 \pm 10$~nm, which compares reasonably well with $\lambda(0) = 232$~nm determined from muon spin rotation measurements \cite{Macdougall2006}. The symbols in Fig.~4 show the data (with the symbol size representing the $\pm 10$ nm error) and the lines show curves expected for single gap BCS superconductors in the clean limit $s-$wave (solid blue), clean limit $d-$wave (dashed green), dirty limit $s-$wave (dot blue) and dirty limit  $d-$wave (dashed - dot green). Clearly, the clean limit weak coupling s-wave curve describes the experimental data almost perfectly in the full temperature range. The inset in Fig.~4 zooms into the low-temperature region showing exponential saturation of the superfluid density approaching $T=0$. Moreover, if we use the measured $\lambda(T) = \lambda(0) + \Delta \lambda(T)$ and fit it to the low - temperature expansion, $\Delta\lambda(T)=\lambda(0)\sqrt{\pi\Delta (0)/2k_BT}\exp{(-\Delta (0)/k_BT)}$, where $\Delta_0$ is the maximum gap value at $T=0$, in the ``low - temperature" range of $T<T_c/3$ we obtain an almost weak - coupling value for the gap to $T_c$ ratio, $\Delta (0)/k_BT_c = 1.71$. The BCS weak - coupling value for isotropic s-wave superconductor is 1.76. Altogether our results convincingly establish MgCNi$_3$ to be a weak - coupling isotropic \textit{s}-wave superconductor.

Next we discuss the measurements of Campbell penetration depth in finite applied DC magnetic field.
Temperature sweeps done in applied fields up to 9 T, where selected curves can be seen in Fig.~5, allow for the construction of the $H-T$ phase diagram, which is shown for two different sample orientations with respect to the applied magnetic field. The single crystal sample of MgCNi$_3$ that was studied was a rectangular bar having approximate dimensions of $0.40\times0.43\times0.73$~mm$^3$.  The two sample orientations about which both the $ac$ and $dc$ magnetic fields were applied are parallel to the long sample axis and perpendicular to the long axis with the fields being also along one of the principle axes of the sample.  Fig.~5 shows that the $H_{c2}$ curves for the single crystalline MgCNi$_3$ are isotropic.  By analyzing these results within the Helfand and Werthamer theory \cite{HW}, we obtain a slope at $T_c$ of -2.63 T/K and $H_{c2}(0)\approx$ 12.3 T.  The values obtained using this analysis are in excellent agreement with those obtained by another group performing resistivity measurements on single crystalline MgCNi$_3$ in applied fields \cite{Lee2008}.

\begin{figure}[b]
\begin{center}
\includegraphics[width=9cm]{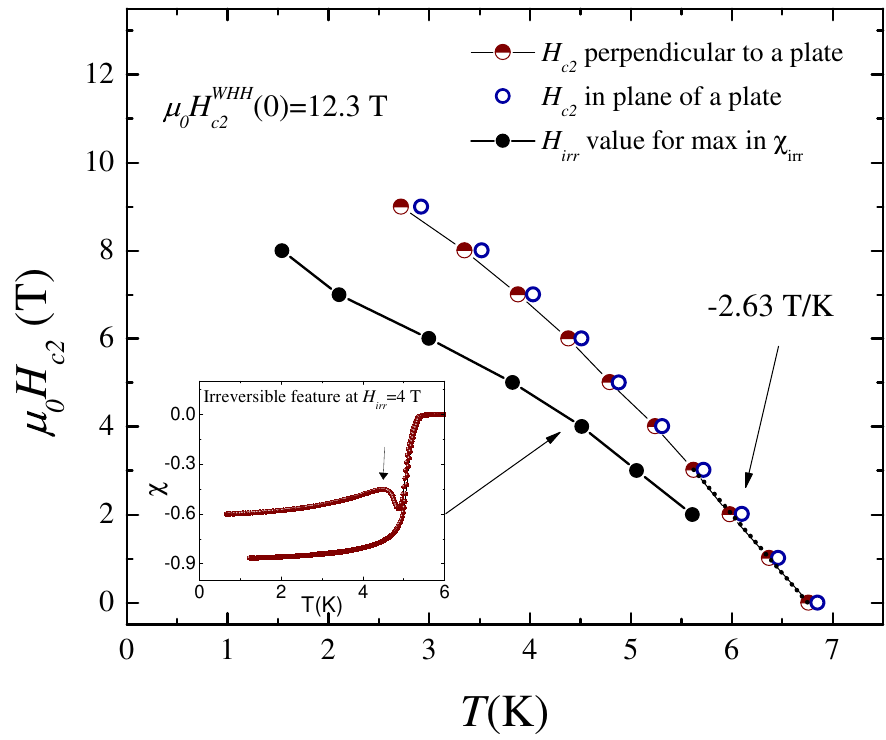}
\caption{(Color online) Field versus temperature diagram the upper critical field for two different crystal orientations as described in the text and also for the location of the maximum in $\chi_{irr}$ (see Fig.~\ref{fishtail}).}
\label{Hc2}
\end{center}
\end{figure}

    Selected runs of TDR frequency shifts vs.\ temperature performed in various applied fields and converted into susceptibility are shown in Fig.~6.  For each run, the sample was cooled in a low field, with the first run cooled in zero field, and then the target field was applied after the sample had been cooled to the base temperature.  The resulting curve is independent of whether or not the sample is cooled in zero field or the previous field run value.  The sample was then field-warmed and field-cooled twice.  Notice from Fig.~6 that the initial zero-field cooled and field-warmed portions of the curve are irreversible, denoted by $\chi_{irr}$.  This irreversibility is believed to be related to a response of the vortex lattice to the applied magnetic field and is a signature of the non - parabolic pinning potential \cite{Prozorov2003_BSCCO}.

\begin{figure}[tb]
\begin{center}
\includegraphics[width=9cm]{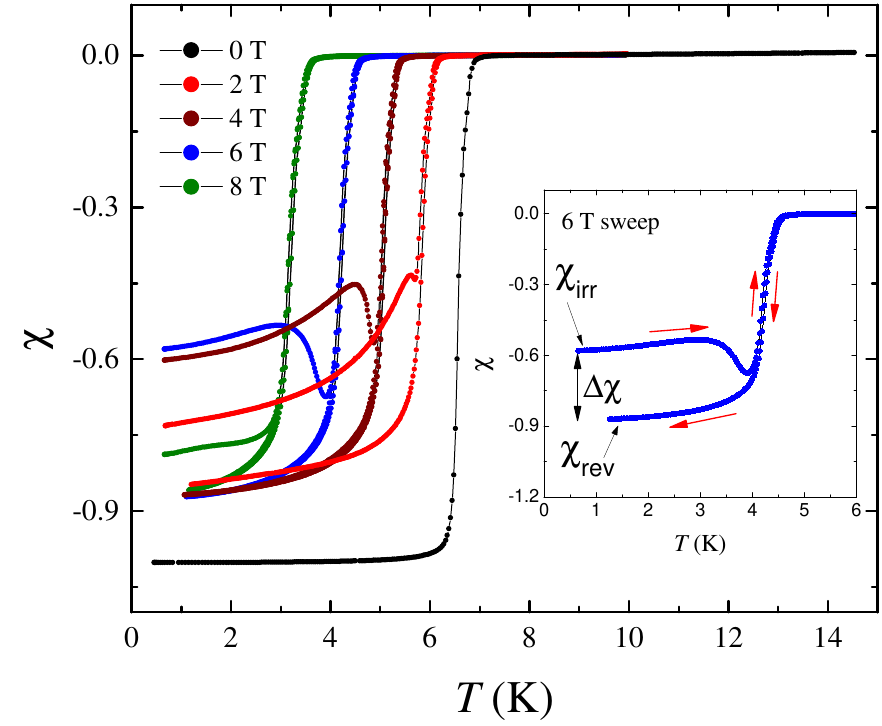}
\caption{(Color online) Temperature scans of the AC magnetic susceptibility of single crystalline MgCNi$_3$ for various values of applied $dc$ magnetic fields up to 8 T.  Each run was zero-field-cooled and then field-warmed and field-cooled twice, which leads to irreversible and reversible portions of the curve.  Inset: The 6 T temperature scan is used here to define the initial irreversible portion of the curve $\chi_{irr}$, the reversible portion obtained after field-warming and cooling $\chi_{rev}$, and the difference $\Delta\chi$ at a temperature of 1.4 K.}
\label{4pichi}
\end{center}
\end{figure}

It should be noted that when considering the magnetic penetration depth of a superconductor in applied $dc$ fields, there are two contributions to the total penetration depth, $\lambda$, when the sample is in the mixed state.  One of these is the usual London penetration depth due to the diamagnetic screening of the applied magnetic field by the condensate, $\lambda_{London}$.  The other component arises from the motion of the vortices and a comprehensive expression has been derived in various works for $\lambda_{vortex}$ \cite{Coffey1991, Brandt1991, vanderBeek1993}.  It has been shown that in the limit of low temperatures and fields that $\lambda_{vortex}$ reduces to the Campbell penetration depth \cite{Campbell1969}, where $\lambda_{Campbell}^2=\phi_0H/\alpha$.  Taking both contributions into account gives the total magnetic penetration depth to be
    \begin{equation}
    \lambda^2=\lambda_{Campbell}^2 + \lambda_{London}^2.
    \end{equation}
    \noindent In the vortex state, the Campbell penetration depth is the dominant term.  Here, $B$ is the applied magnetic field and $\phi_0$ is the flux quantum.  The Campbell penetration depth is important because it contains the necessary information to obtain the Labusch parameter, $\alpha$, which is a measure of the curvature of the potential energy associated with the pinning of vortices.

\begin{figure}[tb]
\begin{center}
\includegraphics[width=9cm]{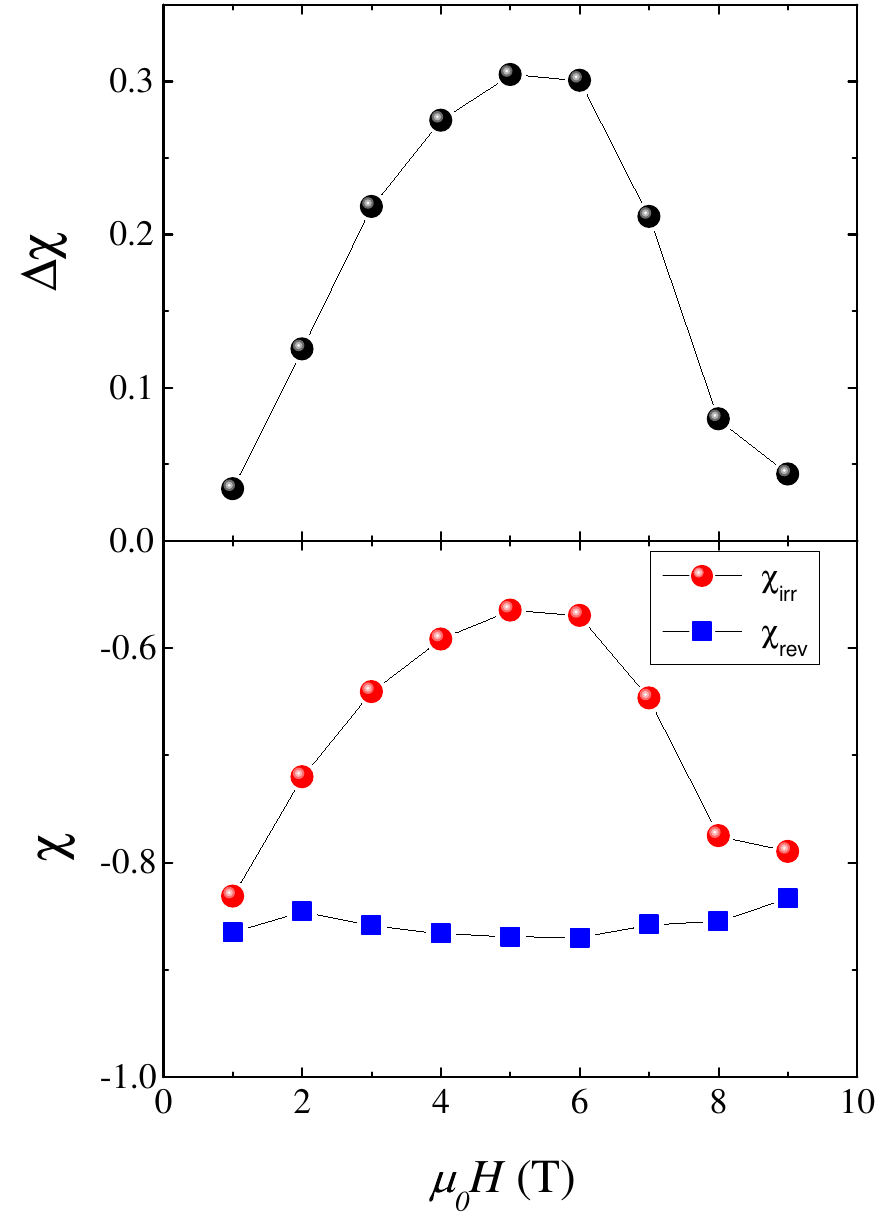}
\caption{(Color online) Upper panel: Susceptibility difference $\Delta\chi=\chi_{irr}-\chi_{rev}$, which can be seen in Fig.~6, at 1.4 K as a function of applied $dc$ magnetic field.  Lower panel: The individual curves of $\chi_{irr}$ and $\chi_{rev}$ taken at 1.4 K as functions of applied $dc$ magnetic field.}
\label{fishtail}
\end{center}
\end{figure}

    The dependence of the susceptibility features $\chi_{irr}$, $\chi_{rev}$ and $\Delta\chi=\chi_{irr}-\chi_{rev}$ (defined in Fig.~6) on applied magnetic field can be seen in Fig.~7.  It is interesting to note the non-monotonic behavior, consistent with a peak in $\Delta\chi$ near 5 T.  This feature, i.e. a maximum in the amount of diamagnetic screening at a particular location in the $H-T$ phase diagram, is commonly referred to as the peak effect \cite{DeSorbo1964} and has been observed in other superconductors like Lu$_2$Fe$_3$Si$_5$ \cite{RTG2008}, MgB$_2$, \cite{Martin2008} the high-$T_c$ BSCCO 2212,\cite{Prozorov2003_BSCCO} and more recently in iron-pnictides \cite{Moll2010, Moll2_2010,Prommapan2011}. Experiments in which the resistivity of clean samples of MgCNi$_3$ with weak pinning have also shown evidence for the existence of the peak effect \cite{Lee2008}.

    Many explanations have been put forth with the intent of explaining the presence of this maximum feature in $\Delta\chi(T,H)$.  One early theoretical work done on the effect of disorder induced pinning on a vortex lattice considered that instead of the usual Abrikosov lattice, there exists a quasi-ordered Bragg glass phase and that the peak effect is a sign of the transition from this phase into a disordered vortex phase \cite{Giamarchi1995}.  TDR experiments on BSCCO 2212 \cite{Prozorov2003_BSCCO} may suggest that the observed hysteresis is a result of ramping the magnetic field after zero-field cooling giving rise to macroscopic screening supercurrents, \textit{j}, which shift the vortices into a state of inhomogeneous distribution, which is in agreement with the critical state (Bean) model.  In this scenario, this procedure gives rise to a state consisting of a displaced vortex lattice, which disappears when the sample is field-cooled due to a relaxation of screening currents.

    In conclusion, good quality single crystals of MgCNi$_3$ were grown at high pressure and studied using DC and AC magnetization.  The zero-field London penetration depth has been measured and converted into the superfluid density, $\rho_s=\left(\lambda(0)/\lambda(T)\right)^2$.  The conventional weak - coupling $s-$wave BCS temperature dependence of the London penetration depth at low temperatures and of $\rho_s(T)$ in the whole temperature range can be reproduced very well from the TDR measurements with a corresponding value of $\lambda(0)$ = 245 nm. The $H-T$ phase diagram has been mapped by measuring $M(T)$ in different applied $dc$ magnetic fields. The $H_{c2}$ is found to be isotropic for two different directions of applied magnetic field with $H_{c2}(0) \approx 12.3$ T by using the standard Helfand and Werthamer analysis. This value corresponds to the coherence length of 5.2 nm and together with  $\lambda(0)$ = 245 nm gives a Ginsburg-Landau parameter of $\kappa \approx 47$.  By studying the effect of field-cooling versus field-warming on the susceptibility, a hysteretic response has been observed and it has been speculated that this arises due a vortex lattice-related phenomenon known commonly as the peak effect and signal non - parabolic nature of the pinning potential.

\section*{acknowledgements}

We thank V. G. Kogan for useful discussions and H. Kim for help with the data analysis. The work at the Ames Laboratory was supported by the U.S. Department of Energy, Office of Basic Energy Sciences, Division of Materials Sciences and Enginnering under contract No. DE-AC02-07CH11358. The work at ETH Zurich was supported by Swiss National Science Foundation, the National Center of Competence in Research MaNEP (Materials with Novel Electronic Properties).

 \end{document}